\documentclass[twocolumn,prl,showpacs,floatfix]{revtex4}
\usepackage{dcolumn}                 
\newcolumntype{w}[1]{D{.}{.}{#1}}
\newcolumntype{.}{D{x}{}{-1}}

\usepackage{times}
\usepackage{amsmath}
\usepackage{amsfonts}
\usepackage{amssymb}
\usepackage{amsthm}
\begin{document}
\preprint{Version 1.1}
\title{Proton charge radius and the perturbative quantum electrodynamics}
\author{Krzysztof Pachucki and Krzysztof A. Meissner}
\affiliation{Faculty of Physics, University of Warsaw,
             Ho\.{z}a 69, 00-681 Warsaw, Poland}
\begin{abstract}
We argue that the proton charge radius conundrum
can be resolved by weakening  the assumption
of perturbative formulation of quantum electrodynamics within the proton.
\end{abstract}
\pacs{12.20.-m, 13.60.-r, 31.30.jr}
\maketitle

The proton radius conundrum refers to the
disagreement between the proton mean square charge radius
determined from muonic hydrogen and from
electron-proton (e-p) systems: atomic hydrogen and
e-p elastic scattering. The muonic hydrogen 
result \cite{pohl, antognini} of $r_p = 0.84087\pm 0.00039$ fm is
about 13 times more precise and much different than the  
CODATA 2010 \cite{codata} result of $r_p = 0.8775\pm0.0051$ fm. 
The CODATA analysis includes atomic hydrogen and 
the precise cross section measurements of Bernauer et
al. \cite{mainz}, which give $r_p = 0.879\pm 0.008$ fm. 
For a recent review, see \cite{review}.

It seems \cite{gilman}, that the most likely 
explanations are related to: (i) novel physics beyond the Standard
Model that differentiates $\mu$-p and e-p 
interactions, (ii) novel two-photon exchange effects that
differentiates $\mu$-p and e-p interactions, (iii) errors
in the e-p experiments. Regarding the Standard Model,
various extensions have been proposed 
to explain this discrepancy, by introducing scalar, 
pseudoscalar or vector particles. The range of available parameters 
is very limited, if not ruled out, mostly due to various precision tests
of the Standard Model, such as $g-2$ of the muon.

Here, we assume that the Standard Model is valid, all the above experiments
are correct, the only incorrect assumption is an implicit one
of the validity of the perturbative QED within the proton.
The elastic scattering off the proton by electrons or muons within the Born
approximation is described by the electric and magnetic formfactors.
Electron interacts with the proton by a single photon exchange, and
what matters is the electromagnetic current of the proton which
can be decomposed into two independent formfactors.
Certainly, the one photon exchange approximation can be supplemented
by the exchange of two-photon, soft photon radiation, photon loops, etc.
However, this requires justification of all assumptions to build quantum 
electrodynamics in the presence of strongly interacting particles.

There is no precise formulation in the literature of Quantum Electrodynamics 
in the presence of hadrons. Let us therefore briefly describe
what we think is generally assumed. It is assumed that it makes sense to 
speak about (chargeless) bare proton and bare neutron, i.e.
hadrons with neither electromagnetic nor weak interactions. 
Such objects would have slightly different masses and different 
electromagnetic formfactors than the real ones.
Now, let us build QED assuming that we have just one proton.
The basic assumption, although usually not spelled explicitly, is that
one can use perturbation theory to account for electromagnetic interactions.
We now want to describe, for example, an electron scattering off a proton using
standard perturbation theory.
The single photon exchange between the electron and 
this bare proton is supplemented by all the proton
dressing diagrams, which change the bare mass and bare formfactors
into physical mass and physical formfactors. The two-photon exchange
diagram will include in addition inelastic contributions.
In such a perturbative picture the electron interacts with the proton
by the photon exchange and this interaction can be fully described by the
electromagnetic current $J^\mu_p$ of the proton. 
However, this has never been verified.
If the perturbative picture of quantum electrodynamics within the proton fails,
one may expect in addition, a different form of an effective electron-proton
interaction. A naive example of such interaction is in 
the elastic positron scattering off an
atomic helium in the ground state. Besides the photon exchange, 
there is the annihilation type of interaction.
In analogy one cannot in principle exclude that the effective electron proton interaction
may contain nonlocal terms beyond the photon exchange, like for example
\begin{equation}
\bar\psi_e(x)\,e^{-i\,e\int_x^y A_\mu\,dz^\mu}\,\psi_e(y)\,
\bar\psi_p(Z)\psi_p(Z)\,F(x-Z,y-Z)
\end{equation}
with implicit insertions of Dirac gamma matrices and with $F$ being some proton formfactor.
Another example can be 
\begin{equation}
J^\mu_p(x)\,j_\mu(x)
\end{equation} 
the product of vector currents with the coupling constant,
that is different for the electron and the muon.
Existence of such interactions is not against the Standard Model
and would explain different charge radius seen by the proton and by the muon, 
because in the nonrelativistic limit they give a local interaction 
of the type $\delta^3(r)$, as the charge radius.
The two (and more) hard photon exchange diagrams lead to additional local
type of interactions but they are much too small to explain
the proton charge radius discrepancy between $e-p$ and $\mu-p$
type of experiments. Can the existence of such additional interactions 
be ruled out on the basis of current experiments?
It seems that not at all. The ratio $G_M/G_E$ of proton formfactors measured
by unpolarized electron-proton scattering is different
from the ratio obtained by polarization transfer electron-proton scattering,
and this difference grows with $Q^2$, square of the momentum exchange.
The difference has been attributed to the two-photon exchange
which contribute more significantly in Rosenbluth separation technique,
used to extract formfactors from unpolarized scattering \cite{venderhagen}.
The evaluation of two-photon exchange diagrams in e-p scattering is problematic,
due to insufficient knowledge of inelastic structure functions 
or, what we would prefer, lack of correct approach to QED on the proton background. 
So the proton charge radius difference can be attributed to the existence of 
additional forms of the effective lepton-proton interaction. 
An agreement for the hyperfine splitting
in H and $\mu$H between theoretical predictions and experimental values,
suggests that these extra forces are spin independent at low $Q^2$,
as the mentioned difference for the $G_M/G_E$ formfactors.
We are not aware of any further experimental verifications of the validity of
perturbative QED within the proton but we see such possibility
by comparison of the electron to positron elastic scattering off the proton.
If there are nonperturbative terms beyond the proton formfactors
the proton charge radius, as seen by positron, can be different from that seen
by the electron. Observation of this difference would be a clear
demonstration that current phenomenological treatment of lepton-hadron
interactions is incomplete.

In summary, if the usual perturbative assumption does not hold, 
the effective interactions of leptons with protons, or more generally hadrons,
may include additional non-standard corrections which apparently may violate 
universality of electromagnetic interactions and explain the observed discrepancies.

\end{document}